\documentclass[aps,prb,twocolumn,longbibliography,groupedaddress,superscriptaddress]{revtex4-2}
\usepackage[normalem]{ulem}
\usepackage{graphicx}  
\usepackage{dcolumn}   
\usepackage[english]{babel}
\usepackage{bm}        
\usepackage{amsmath, amsthm, amssymb}
\usepackage{bbold} 
\usepackage{mathrsfs}
\usepackage{array}
\usepackage[usenames]{color}
\usepackage{soul} 
\usepackage{ulem}
\usepackage{chemformula}
\usepackage{hyperref}
\usepackage{float}
\usepackage{changes}

\emergencystretch=\maxdimen
\definecolor{royalblue}{RGB}{4,51,255}
\definecolor{eggplant}{RGB}{180,33,147}
\definecolor{darkgreen}{RGB}{0,102,51}

\newcommand{\ii}{\ensuremath{\mathbf{i}}}
\newcommand{\jj}{\ensuremath{\mathbf{j}}}
\newcommand{\Z}{\ensuremath{\mathcal{Z}}}
\newcommand{\bra}[1]{\langle #1 |}
\newcommand{\ket}[1]{| #1 \rangle}

\begin{document}

\title{Bose-Einstein condensation in honeycomb dimer magnets and \ch{Yb2Si2O7}}
\author{Chunhan Feng}
\affiliation{Center for Computational Quantum Physics, Flatiron Institute, New York, New York 10010, USA}
\affiliation{Department of Physics, University of California, Davis, CA 95616, USA}
\author{E. Miles Stoudenmire}
\affiliation{Center for Computational Quantum Physics, Flatiron Institute, New York, New York 10010, USA}
\author{Alexander Wietek}
\affiliation{Max Planck Institute for the Physics of Complex Systems, N\"othnitzer Strasse 38, Dresden 01187, Germany}
\affiliation{Center for Computational Quantum Physics, Flatiron Institute, New York, New York 10010, USA}

\date{\today}

\begin{abstract}
An asymmetric Bose-Einstein condensation (BEC) dome was observed in a recent experiment on the quantum dimer magnet $\rm Yb_2Si_2O_7$, which is modeled by a ``breathing" honeycomb lattice Heisenberg model with possible anisotropies. We report a remarkable agreement between key experimental features and predictions from numerical simulations of the magnetic model. Both critical fields, as well as critical temperatures of the BEC dome, can be accurately captured, as well as the occurrence of two regimes inside the BEC phase. Furthermore, we investigate the role of anisotropies in the exchange coupling and the $g$-tensor. While we confirm a previous proposal that anisotropy can induce a zero temperature phase transition at magnetic fields smaller than the fully polarizing field strength, we find that this effect becomes negligible at temperatures above the anisotropy scale. Instead, the two regimes inside the BEC dome are found to be due to a non-linear magnetization behavior of the isotropic breathing honeycomb Heisenberg antiferromagnet. Our analysis is performed by combining the density matrix renormalization group (DMRG) method with the finite-temperature techniques of minimally entangled typical thermal states (METTS) and quantum Monte Carlo (QMC).

\end{abstract}

\maketitle

\noindent

\section{Introduction} Quantum magnets exhibit many phenomena which currently elude our understanding~\cite{Lacroix2011,Zapf2014,Savary2016}. 
The combination of quantum and thermal fluctuations of local magnetic moments 
combined with possible geometric frustration can lead to the emergence
of entirely new states of matter. 
As computational methods for quantum many body
systems have significantly advanced in recent years~\cite{LeBlanc2015,Cirac2021,Wietek2021b}, 
the bridge between experimental observations and  explanations using theoretical models 
can increasingly be built not only on a qualitative, but also quantitative level. 
 
A particularly interesting phenomenon in quantum magnetism is the Bose-Einstein
condensation of triplons in quantum dimer magnets \cite{Zapf2014}. A broad variety of compounds have to date been found to exhibit this magnetic
analogue of superfluidity in \ch{^{4}He}, including \ch{BaCuSi2O6}~\cite{Sasago1997, Jaime2004,Ruegg2007},  \ch{TlCuCl3}~\cite{Tanaka2001,Ruegg2003,Yamada2007} and 
\ch{Ba3Cr2O8}~\cite{Nakajima2006,Aczel2009,Kofu2009}. 
Here the magnetic field acts as the chemical
potential condensing the bosonic triplons, which are the elementary excitations
of the local spin singlet dimers. This condensation at a critical value of the 
magnetic field $H_{c_1}$ causes the system to order antiferromagnetically at low temperatures, before
reaching a fully spin-polarized state beyond a larger critical magnetic field
$H_{c_2}$. The intervening antiferromagnetic or BEC phase forms a dome in the temperature versus 
field phase diagram, the maximum temperature of this dome ranging from a few hundred milli-Kelvin to around $10$ Kelvin 
depending on the compound. Typical values of $H_{c_1}$ and $H_{c_2}$ range from $1$ to $100$ Tesla.

Conventionally, the BEC dome constitutes a single phase of matter. But surprisingly, 
recent experimental results on the material \ch{Yb2Si2O7}~\cite{Hester2019}, discovered
two distinct regimes separated by another field value $H_m$ between the $H_{c_1}$ and
$H_{c_2}$ fields inside the BEC dome of this compound. The critical magnetic fields $H_{c_1} \approx 0.4 \rm T$ and $H_{c_2} \approx 1.2 \rm T$ have been found to be low compared to similar quantum dimer magnets. The two regimes are distinguished by a change in the field dependence of the magnetization and the related ultrasound velocity. Moreover, while the regime at smaller magnetic fields features a sharp anomaly in the specific heat which is absent in the larger field regime. The magnetic properties of this compound can be modeled by a ``breathing" honeycomb antiferromagnet. To explain the peculiar features observed in experiment, a recent work has highlighted the role of anisotropies in both the spin-exchange as well as the $g$-tensor~\cite{Flynn2021}. Here, we extend this analysis to study the full finite-temperature phase diagram of the model proposed in Ref.~\cite{Flynn2021}. We establish critical fields and temperatures of the BEC phase as well as the existence of two regimes inside the BEC dome signalled by a change in the behaviour of the magnetization process. A particular focus will be on the role of possible anisotropies at finite temperature.

 \begin{figure}[!htb]
  \includegraphics[width=1\columnwidth]{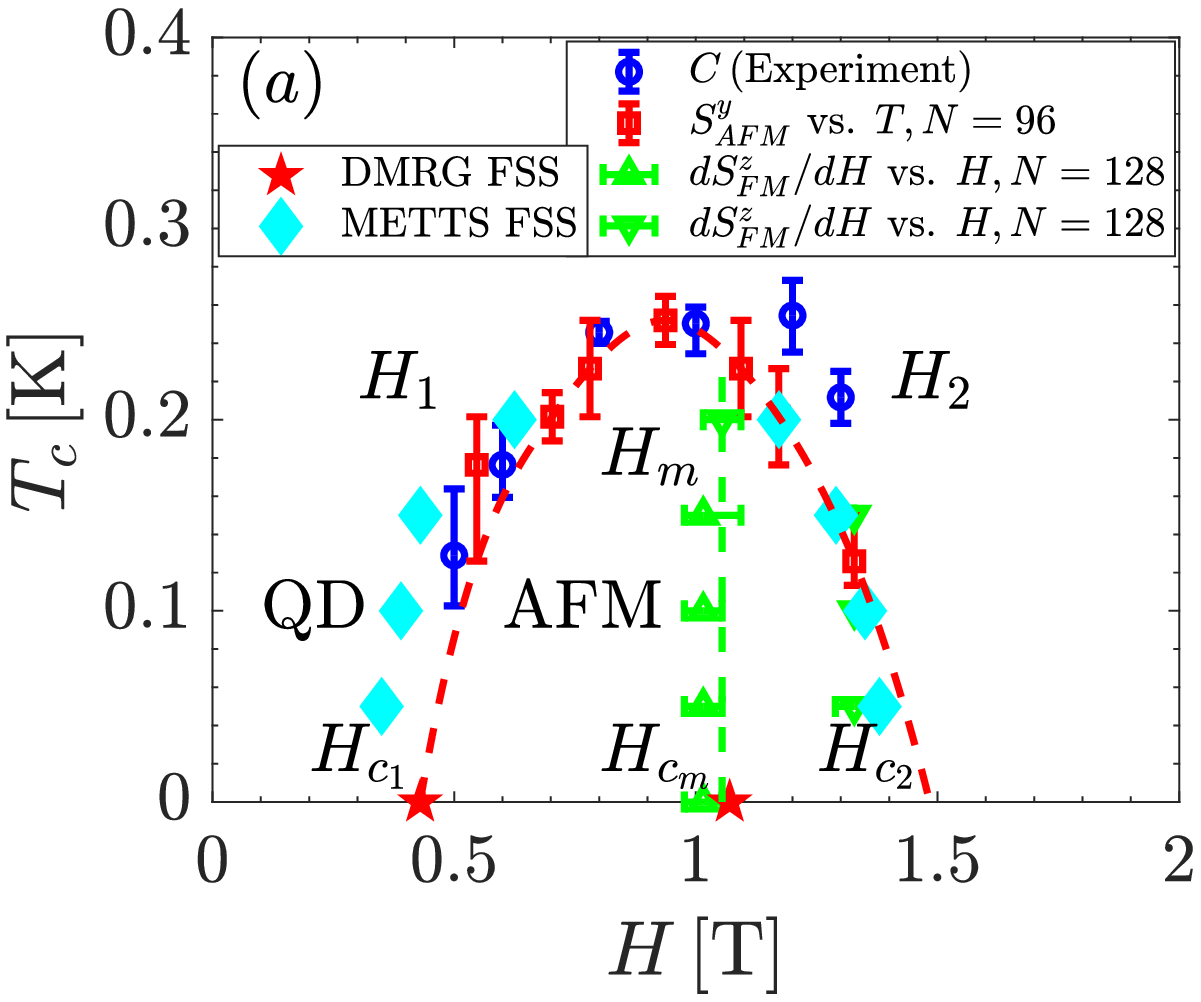}
  \includegraphics[width=1\columnwidth]{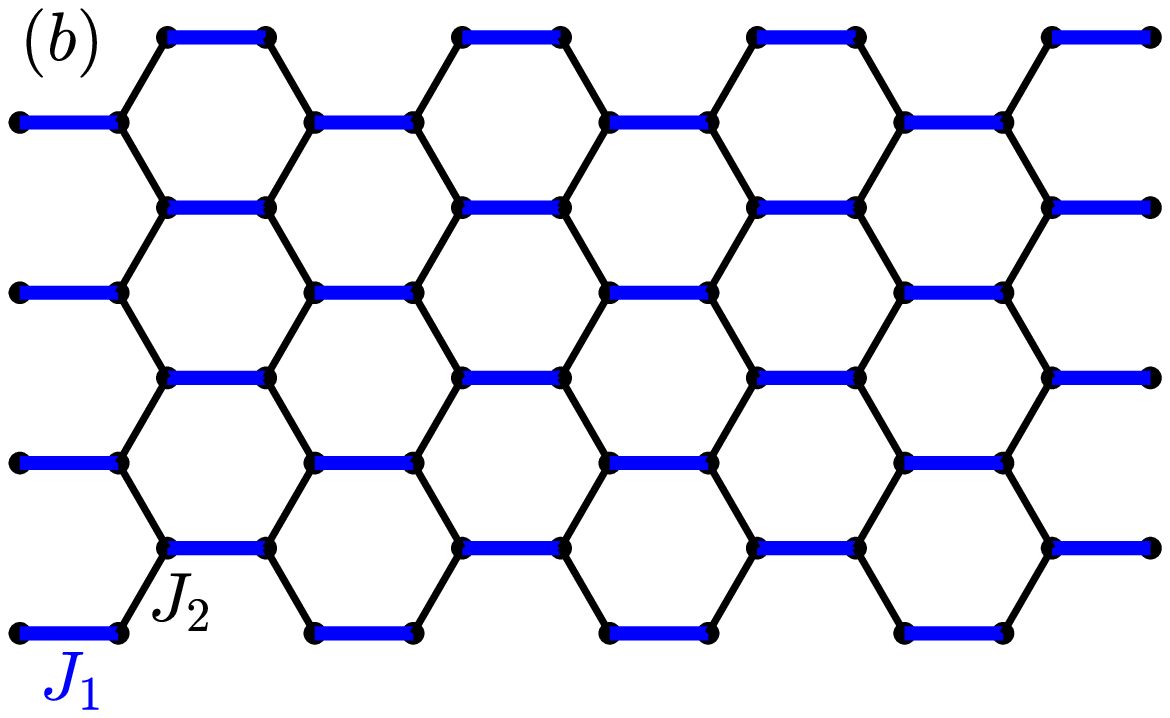}
 \caption{(a) Phase diagram: critical temperature $T_c$ (Kelvin) versus magnetic field $H$ (Tesla). The blue circles are the peak positions of heat capacity $C$ versus temperature $T$ from the experiment \cite{Hester2019}, while red squares are obtained by locating the largest slope points in $S_{AFM}^y$ versus $T$ for a $\rm Z4 \_12$ system with $N=96$ sites. Green triangles (the vertical dashed line inside the dome) indicate the slope changes in $dS_{FM}^z/dH$ on a $\rm Z4 \_16$ honeycomb lattice with $N=128$ sites, related to an analogous shift in the ferromagnetic Bragg peak and the ultrasound velocity in experiment. The red dashed curve is only a guide to the eye. A finite size scaling analysis is also conducted to obtain the crossings, which are revealed by red stars (for the ground state using DMRG data) and blue diamonds (for finite temperatures using METTS data). (b) Geometry of a ZW\_L simulation cylinder with $W=4$ and $L=8$. $W$ refers to the width (y-direction size) of the cylinder and $L$ to the length (x-direction size). We use periodic boundary condition (PBC) for y direction and open boundary condition (OBC) for x direction. 
 }
 \label{fig: phase_diagram}
 \end{figure}
\section{Model and Methodology}

 We study a “breathing" Heisenberg antiferromagnetic with additional anisotropies model previously proposed in Refs.~\cite{Hester2019,Flynn2021}, given by
\begin{align} 
\mathcal{H} =  \sum_{\langle \ii, \jj \rangle, \alpha} J_{\ii \jj}^{\alpha}\ S_{\ii}^{\alpha} S_{\jj}^{\alpha} - H \sum_{\ii, \alpha} g_{z\alpha} S_{\ii}^{\alpha} ,
\label{eq:ham}
\end{align}
where $\ii, \jj$ are lattice sites, $\langle \ii,\jj \rangle $ denotes the nearest neighbors on a honeycomb lattice and $\alpha=x,y,z$ indicates the spin directions. A breathing honeycomb lattice is shown in Fig.~\ref{fig: phase_diagram}(a). The couplings $J_1$ on blue horizontal bonds are stronger than couplings $J_2$ on the remaining nearest neighbor bonds, $J_1 > J_2$. In the limit $J_2 =0$, the ground state is a product of local singlets on the dimers. In experiments~\cite{Hester2019}, the coupling constants have been estimated to be $J_1=0.217(3) \rm meV$ and $J_2=0.089(1) \rm meV$, corresponding to a ratio of $J_2 / J_1=0.41003221$. 
Similar to Ref.~\cite{Flynn2021}, we consider a spin anisotropy by setting $J_{ij}^y=(1+\lambda)J_{ij}^x$, $J_{ij}^z=J_{ij}^x=J_1, \rm{ resp. } J_2$, and $\lambda=0.03$. Such a small $\lambda$ guarantees the physics can be mainly characterized by the Heisenberg model. This spin anisotropy breaks the spin SU($2$) rotations symmetry down to a remaining U($1$) symmetry with the $Y$ axis as a principal axis of rotation. A further anisotropy is introduced by a non-isotropic $g$-tensor. We consider $g_{zy}=0$, and a staggered coupling in $X$ direction is used $|g_{zx}|=g_{zz}/500$, $g_{zx}^A=-g_{zx}^B$ for sublattice A and B.
This additional anisotropy further breaks down the U($1$) symmetry to a remaining $\mathbf{Z}_2$ symmetry for $H\neq 0$. To emphasize how closely the experiments on \ch{Yb2Si2O7} are captured by our results, all results in this manuscript are reported in experimental units, set by $J_1=0.217(3) \rm meV$,  $J_2=0.089(1) \rm meV$, and $g_{zz} = 4.8$. 

To study the system with Hamiltonian Eq.~(\ref{eq:ham}) we use three numerical methods. For zero-temperature properties we use the DMRG algorithm \cite{PhysRevLett.69.2863,SchollwoeckMPS}. For properties at a finite temperature $T$, we use the minimally entangled typical thermal states (METTS) approach \cite{white2009minimally,stoudenmire2010minimally,Wietek_Stripes,Wietek2021b,Feng2022}. Both these methods are implemented using the ITensor software (C++ version) \cite{ITensor}.

The METTS algorithm samples a set of quantum states whose average yields controlled finite temperature results. Unlike quantum Monte Carlo methods, METTS does not encounter sign or complex phase problems that would occur in our model Eq.~(\ref{eq:ham}) from the magnetic field term 
coupling to multiple spin components.
The METTS algorithm is motivated as follows: the expectation value of an observable $\mathcal{O}$ can be expressed as 
\begin{align} 
\nonumber \mathcal{\langle O \rangle_\beta} 
&=  \frac{1}{\Z} \text{Tr} \big[ e^{-\beta H}  \mathcal{O} \big] \\
&=  \frac{1}{\Z} \sum_{i} \bra{i} e^{-\beta H/2} \mathcal{O} e^{-\beta H/2} \ket{i} \\
&= \frac{1}{\Z} \sum_{i} P(i)\, \bra{\phi(i)} \mathcal{O} \ket{\phi(i)}
~~,
\label{eq:expectation}
\end{align}
where 
\begin{align}
|\phi(i)\rangle & = P(i)^{-1/2}\ e^{-\beta H/2} \ket{i} \\ 
P(i) & = \bra{i} e^{-\beta H} \ket{i} \ .
\end{align}
Here $\Z$ is the partition function and $\ket{i}$ is an orthonormal basis of classical product states. The states $|\phi(i)\rangle$ are known as METTS. To calculate $|\phi(i)\rangle$, we use matrix product states (MPS) to evolve the states $|i\rangle$ in imaginary time, using a combination of Trotter gates and the TDVP algorithm to perform the time evolution \cite{Wietek_Stripes,TDVP}. We also take advantage of the METTS pure state algorithm in our simulations, constructing the next METTS from a product state obtained by collapsing the previous METTS, which guarantees quantum states are sampled efficiently with the desired distribution. The maximum bond dimension required to use a MPS to represent the state increases exponentially in the width of two-dimensional lattices hence we restrict the width of the honeycomb lattice to be 4 in this work. 
Finally, for studying the isotropic case where $\lambda = 0$ and $g_{zx} = g_{xz} = 0$ we employ quantum Monte Carlo simulations in the form of the worm algorithm ~\cite{Suwa2010,Todo2022}.



\section{Phase Diagram}
The main result of this paper is a temperature $T_c\ [\rm {K}]$ versus magnetic field $H\ [\rm{T}]$ phase diagram shown in Fig.~\ref{fig: phase_diagram}. We will use $H_{c_1}$, $H_{c_m}$, $H_{c_2}$ to represent the critical magnetic fields in the ground state and $H_1$, $H_m$, $H_2$ at finite temperatures. We determine the phase boundaries several different ways: (1) We conduct a finite-size scaling analysis for the antiferromagnetic (AFM) structure factor $S_{AFM}^y$ for spin $y$  in the ground state given by DMRG (red stars, $H_{c_1}$ and $H_{c_m}$) and at finite temperatures obtained by METTS (blue diamonds, $H_1$ and $H_2$). We use the  crossings of rescaled $S_{AFM}^y$ for different system sizes to locate the transitions. 
(2) For different magnetic fields $H$, we locate the temperature $T^*$ (red squares) at which $S_{AFM}^y$ versus $T$ curves have the largest slope. 
(3) We compute the derivative of the ferromagnetic structure factor for the spin $z$ component with respect to magnetic field, $dS_{FM}^z/dH$, which is the derivative of the magnetic Bragg peak intensity (proportional to the square of net magnetization) with respect to magnetic field in the experiment, which behaves analogously to the ultrasound velocity \cite{Hester2019}. Magnetic field values $H_m$ where the slope of the curves changes significantly are shown as green triangles along the vertical dashed line in Fig.~\ref{fig: phase_diagram}. The peak positions $H_2$ in $dS_{FM}^z/dH$ v.s. $H$ are denoted by green down triangles along the right boundary of the dome. 

The phase diagram given by the heat capacity in the experiment is also shown as blue circles in Fig.~\ref{fig: phase_diagram} as a comparison to our simulations. All these approaches of obtaining the transition points will be discussed in more detail in the following sections. 

\section{Ground State Properties}
\label{sec:Ground State Properties}
\begin{figure}[!htb]
 \includegraphics[width=1\columnwidth]{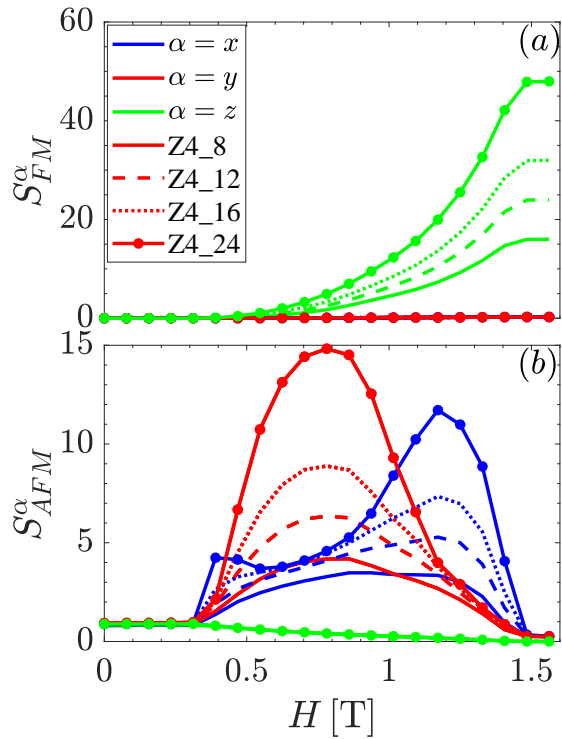}
 \caption{Ferromagnetic and anti-ferromagnetic structure factors $S_{FM}, S_{AFM}$versus magnetic field $H$ for different spin $\sigma=x,y,z$  components. Different line types represent the results for different lattice sizes }
 \label{DMRG}
 \end{figure}
 \begin{figure}[!htb]
 \includegraphics[width=1\columnwidth]{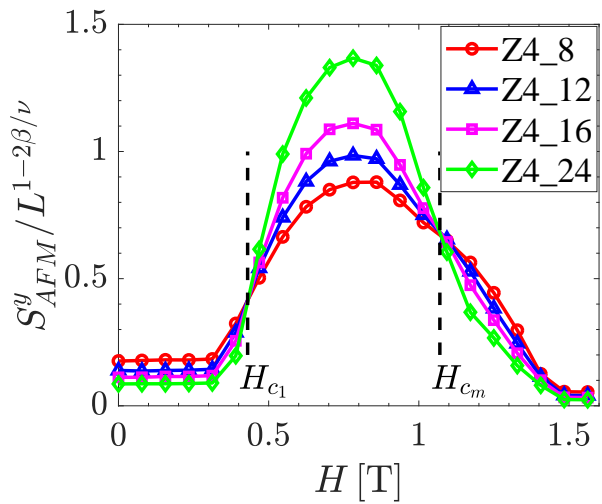}
 \caption{Finite-size scaling analysis for anti-ferromagnetic structure factor (spin y component) $S^y_{AFM}$. $\beta=1/8, \nu=1$ are 2D Ising critical exponents. The two crossing points in the plot indicate $H_{c_1} \sim 0.43 \, \rm T $ and $H_{c_m}\sim 1.07 \, \rm T $ }
 \label{fig:DMRG_FSS}
 \end{figure}
We perform DMRG calculations to study the ground state physics of the system and verified the results presented in \cite{Flynn2021}. We investigate the magnetic structure factor,

\begin{align} 
S^\alpha(\mathbf{q}) = \frac{1}{N} \sum_{ \ii, \jj  } \langle S_{\ii}^\alpha S_{\jj}^\alpha \rangle e^{i\mathbf{q} \cdot (\ii -\jj)},
\label{eq:structure_factor}
\end{align}
where N is the total number of sites and $\mathbf{q}$ denotes the momentum. Results on various system sizes are shown in Fig.~\ref{DMRG}. We investigate both the ferromagnetic (FM) and anti-ferromagnetic (AFM) structure factors, $S^{\alpha}_{FM}=S^{\alpha}(\mathbf{q}=(0,0))$ and $S^{\alpha}_{AFM}=S^{\alpha}(\mathbf{q}=M)$.
When the external magnetic field is relatively small, $H \lesssim H_{c_1} \sim 0.43 \, \rm T$, the ground state is in the singlet quantum dimer phase, adiabatically connected to a product state of singlet dimers. Thus, $S^\alpha_{FM}$ vanishes but $S^\alpha_{AFM}$ retains a finite value. In the middle of the dome, $S^y_{AFM}$ increases to its maximum at around $H \sim 0.8 \, \rm T$ and then decreases. In the ordered AFM phase, $S^y_{AFM}$ is roughly proportional to the lattice size $N$, indicating a long-range order has developed. When $H \gtrsim 1.5 \, \rm T$, the ground state is the spin polarized phase. $S^\alpha_{AFM}$ vanishes and $S^z_{FM}$ retains a finite value.  

In order to locate the transition magnetic field $H_{c_1}$ and $H_{c_m}$ more accurately, we conduct a finite-size scaling analysis. Since a nonzero $g_{zx}=1/500$ introduces a tiny staggered magnetic field in spin $x$ direction, the AFM pattern in the $x$ spin component is a consequence of the field rather than a spontaneous symmetry breaking. Therefore to investigate spontaneous symmetry breaking and phase transitions we focus only on AFM order in the spin $y$ direction.

When $H_{c_1} < h < H_{c_m}$ the ground state breaks a $\mathbb{Z}_2$ symmetry. Thus, we may expect that the magnetic transition exhibits universal finite-size scaling of $S^{y}_{AFM}$ described by the 2D Ising critical exponents $\beta=1/8$ and $\nu=1$. The re-scaled AFM structure factor $S^y_{AFM}/L^{1-2\beta/\nu}$ is plotted as a function of magnetic field $H$ in Fig.~\ref{fig:DMRG_FSS}. The two crossing points appearing in the plot indicate $H_{c_1} \sim 0.43 \, \rm T $ and $H_{c_m} \sim 1.07 \, \rm T$, which is in agreement with \cite{Flynn2021}. When $H_{c_1} \lesssim H \lesssim H_{c_m}$, AFM order appears in both spin $x$ and $y$, while a long range AFM order in spin y vanishes and the order in x dominates when $H_{c_m} \lesssim H \lesssim H_{c_2}$.

\section{Finite Temperature Properties} 
\label{sec:finite temperature properties}
\begin{figure}[!htb]
 \includegraphics[width=1\columnwidth]{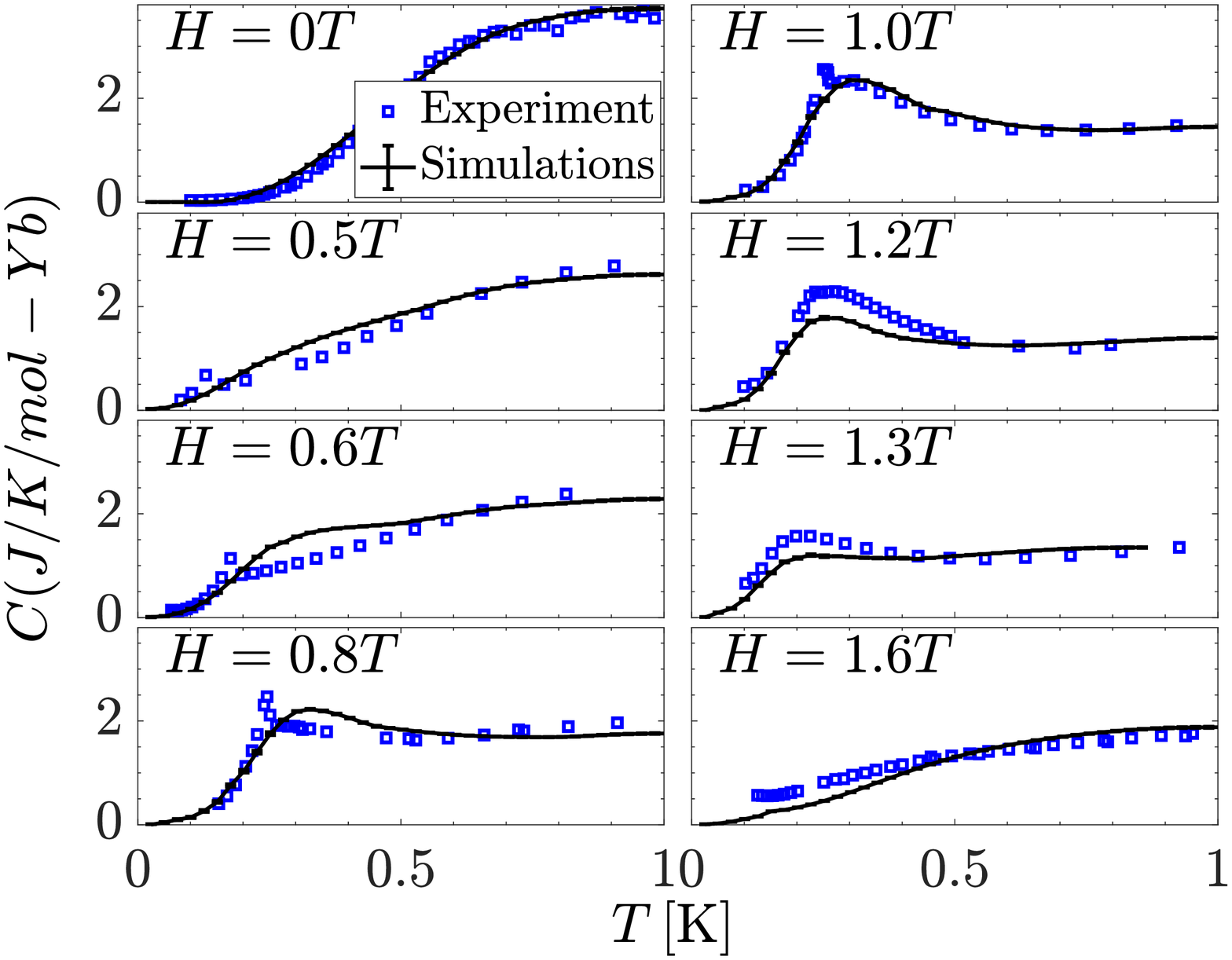}
 \caption{Heat capacity $C$ as a function of temperature $T$ for several different magnetic field $H$ on the Z4\_12. At $H=0 \, \rm T$, the system is in the quantum dimer phase and heat capacity displays a broad Schottky peak at $T \sim 1 \, \rm K$. The system transits to AFM phase when $0.5 \, \rm T \lesssim H \lesssim  \, \rm 1.0 \, \rm T$, and a sharp peak is observed. In this region, the transition temperature $T_c$ increases as magnetic field goes up. When magnetic field further increases, $H \sim 1.2-1.3 \, \rm T$, $T_c$ descends with $H$ increasing, tracking the right hand side of the BEC dome. When $H \sim 1.6 \, \rm T$ or higher, the broad peak shifts to higher temperatures in the polarized phase as expected.}
 \label{fig:C_T}
 \end{figure}
After verifying the ground state phase diagram, we move to finite temperature properties, which are the main focus of this paper. The heat capacity $C=dE/dT \, \rm [J/K/(mol-Yb)]$ versus temperature $T \, \rm [K]$ for several different magnetic fields is shown in Fig.~\ref{fig:C_T}. The simulation results (blue dots) match with the experiment data (solid black curve) reasonably well, especially considering the limited system sizes used in the METTS calculations. In the absence of magnetic field, specific heat exhibits a broad maximum at $\sim 1 \, \rm K$. When $0.5 \, \rm T \lesssim H \lesssim  \, \rm 1.0 T$ a sharp anomaly is observed in the experiment indicating a transition to a long range AF order existing in the system, which will be further verified by investigating magnetic structure factors. The transition temperature $T_c$ goes up as the magnetic field $H$ increases from $0.5 \, \rm T$ to $1.0 \, \rm T$. This maps out the left boundary of the BEC dome in the H vs T phase diagram. Although the peaks in heat capacity curves given by simulations are not as sharp as those in experiments, possibly due to the finite lattice size effect or other type of interactions in real materials which cannot be fully characterised by the Hamiltonian, the positions of the peaks and the main feature of the curves can still be reflected by simulations. 
 \begin{figure}[!htb]
 \includegraphics[width=1\columnwidth]{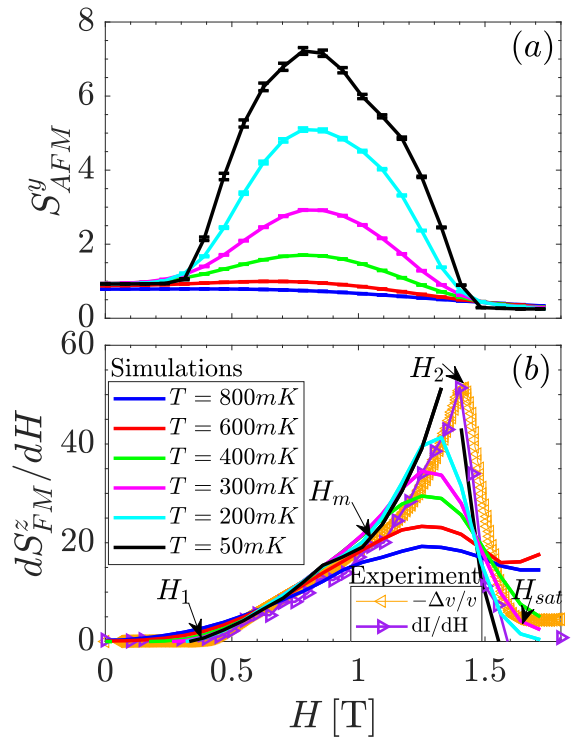}
\caption{(a) Anti-ferromagnetic structure factor for spin Y, $S_{AFM}^{y}$ as a function of magnetic field $H$ (Tesla) (b) The derivative of ferromagnetic structure factor for spin Z $S_{FM}^{z}$ as a function of magnetic field $H$ (Tesla).
The inverse of the ultrasound velocity $-\Delta v/v$ and the derivative of the Bragg peak intensity $dI/dH$ at $T=50 \, \rm mK$, obtained in experiment \cite{Hester2019} are shown in orange and purple lines respectively as comparisons. The peak value of these experiment data are rescaled to match the maximum of our simulation results.
}
 \label{fig:SF_H}
 \end{figure}

When $H$ further grows up to $0.8 \, \rm T$, a broad feature is noticed in the curve and dominates above $H_{c_m}$. The location of the heat capacity peak moves to lower temperatures when the magnetic field further increases. This corresponds to the phase boundary $H_2$ of the high magnetic field region of the dome. It describes a transition from AF order to a fully polarized paramagnetic phase. As expected, in the paramagnetic phase, $H\gtrsim H_{c_2} \sim 1.4 \, \rm T$ the broad peak shifts to higher temperatures as magnetic field increases.    

Similar to our previous ground state study, we explore the FM and AFM structure factors for finite temperatures as well. The spin $y$ antiferromagnetic structure factor $S_{AFM}^y$ is examined as a function of the magnetic field $H$ in Fig.~\ref{fig:SF_H}(a). In the quantum dimer  or singlet phase, $H \lesssim H_1 \sim 0.43 \, \rm T$, $S_{AFM}^y$ has a finite value. When $H_1 \lesssim H \lesssim H_2$, a dome appears in $S_{AFM}^y \, \rm versus \, H$, indicating the AFM order in spin Y develops. The magnitude of the structure factor goes up as temperature decreases as expected. When the magnetic field becomes relatively large, spins tend to be in the same direction as the magnetic field. 
The FM order in the spin Z component is observed and $S_{AFM}^y$ almost vanishes as $H \gtrsim H_2$. We take the derivative of $S_{AFM}^y$ with respect to $H$ to locate the largest slope position $H^*$ and mark them by red squares in the phase diagram Fig.~\ref{fig: phase_diagram}. Although this cannot be viewed as an accurate method to determine the phase transition points, it can give us a rough estimate of the phase boundary of the BEC dome, inside which AFM order develops.

In addition, the derivative of ferromagnetic structure factor in spin Z with respect to magnetic field $dS_{FM}^z/dH$, is plotted as a function of $H$ in Fig.~\ref{fig:SF_H}(b). It is the derivative of Bragg peak intensity, which behaves similar to the ultrasound velocity in the experiment. $dS_{FM}^z/dH$ is almost $0$ when 
\mbox{$H\lesssim H_{c_1} \sim 0.43 \, \rm T$}. A significant slope change occurs at $H_{m} \sim 1.07 \, \rm T$ at low temperatures and we denote these points by green triangles in Fig.~\ref{fig: phase_diagram}. 
This change in slope was highlighted in experiment as a main indication of the occurrence of two regimes.
The positions of the peak correspond to $H_{2}\sim 1.4 \, \rm T$, denoted by green down triangles in Fig.~\ref{fig: phase_diagram}. When $H \gtrsim 1.6 \, \rm T$, $S_{FM}^z$ saturates and hence $dS_{FM}^z/dH$ goes down to and stays at a value close to $0$ at low temperatures.
\begin{figure}[!htb]
 \includegraphics[width=1\columnwidth]{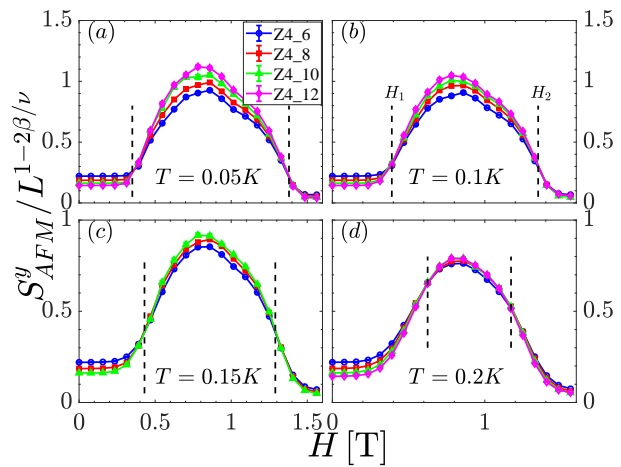}
 \caption{Finite size scaling analysis: rescaled anti-ferromagnetic structure factor $S_{AFM}^y/L^{1-2\beta/\nu}$ versus magnetic field $H$ [T] for several different temperatures. $2$D Ising critical exponents $\beta=1/8, \nu=1$ are used. The two crossings in each plot at different temperatures $T$, $H_1$(left) and $H_2$(right) are denoted by cyan diamonds in Fig.~\ref{fig: phase_diagram}.}
 \label{FSS_finiteT}
 \end{figure}
 
In order to give a more accurate value for the transition out of the ordered phase at larger fields, we apply finite-size scaling analysis to the system at several temperatures, $T=0.05, 0.1, 0.15, 0.2 \, \rm K$. Since the order breaks a $Z_2$ symmetry, we fit to a scaling form using the $2D$ Ising critical exponents $\beta=1/8, \nu=1$. Two crossing points $H_1$ and $H_2$ are observed as expected. We present these data points by cyan diamonds in Fig.~\ref{fig: phase_diagram}. These points basically follow the boundary of the BEC dome instead of locating at the green dashed line around $H_{c_m}\sim 1.07 \, \rm T$. 
This suggests the  ``intermediate" order in the ground state found in \cite{Flynn2021}---which was attributed to a non-zero value of $g_{zx}$--is in fact quickly washed out as temperature $T$ increases such that the transition point likely moves rapidly from $H_{c_m}\sim 1.07 \, \rm T$ at $T=0 \, \rm K$ (red star) to $H_2 \sim 1.4 \, \rm T $ at $T=0.05 \, \rm K$ (cyan diamond). Although this ``intermediate phase" therefore disappears at finite $T$, the slope changes in $dS_{FM}^z/dH$ discussed above can still be observed, implying the similar phenomena observed in ultrasound velocity in the experiment at $H_{c_m}$ is more likely due to a crossover rather than a phase transition. More evidence to support this argument will be shown in the texts and plots below, where we will see that it is a very general feature and rather insensitive to details such as the value of $g_{zx}$.

\section{The isotropic model}
As is discussed in Sec. \ref{sec:finite temperature properties}, the critical magnetic field indicated by the second crossing point in the finite-size scaling analysis moves quickly from $H_{c_m} \sim 1.07 \, \rm T$ in the ground state to $H_2 \sim 1.4 \, \rm T$ at a small temperature $T=0.05 \, \rm K$, implying the tiny staggered magnetic field in the spin $x$ direction (small $g_{zx}$ value) might not be the correct explanation of the feature at $H_{c_m}$ shown as a vertical green dashed line in Fig. \ref{fig: phase_diagram}. Hence, we employ simulations for $g_{zx}=0$ in this section and compare them with results in Sec. \ref{sec:finite temperature properties}, \ref{sec:Ground State Properties} with $g_{zx}=1/500$ to explore the true effects of $g_{zx}$.

First we do a ground state ($T=0$) finite-size scaling analysis for the same model without a staggered field in spin $x$, i.e. $g_{zx}=0$. As expected, the crossing point occurs at $H \sim 1.5 \, \rm T$ (close to $H_{c_2}$) instead of happening at $H_{c_m} \sim 1.07 \, \rm T$ in the original model with $g_{zx}=1/500$. Thus it coincides with the bottom-right edge of the dome computed from our finite-temperature calculations.

Next we examine $dS_{FM}^z/dH$ for $g_{zx}=0$. The slope changes observed in Fig.~\ref{fig:SF_H} appear quite similar to the $g_{zx}=0$ case, and are again analogous to the slope changes in the ultrasound velocity signaling the vertical phase boundary in the experiment \cite{Hester2019}. All the evidence above implies that the model with a tiny staggered magnetic field  in spin X direction (non-zero $g_{zx}$) can give rise to an intermediate phase only at $T=0$ and very small values of $T$. Such a model is therefore not able to explain the vertical $H_m$ line observed in the experiment within the interpretation of this line being a true phase transition.
The vertical phase boundary suggested by the ultrasound velocity in the experiment (or the derivative of FM structure factor) is more likely to be a crossover instead. 

Finally, we investigated the fully isotropic model, where both $g_{zx}=0$ and $\lambda=0$. In this case, QMC can be applied without encountering a sign problem~\cite{Todo2022}. $dS_{FM}^z/dH$ is shown for the fully isotropic case on a $16 \times 16 \times 2$ lattice in Fig.~\ref{fig:SF_H}(c). Inside the BEC dome we indeed also observe two different regimes. For $0.5 \rm{T} \lesssim H \lesssim 1.0 \rm{T}$, $dS_{FM}^z/dH$  is only weakly dependent on temperature, while for $1.0 \rm{T} \lesssim H \lesssim 1.5 \rm{T}$ we observe that the peak close to the saturation field only develops at lower temperatures. This behavior is exactly what is observed in experiments by measuring the Bragg peaks and ultrasound velocity. Thus, the occurrence of two regimes in the magnetization process is intrinsic to the isotropic breathing honeycomb antiferromagnet and not necessarily related to an anisotropy in either the spin-spin interactions or the coupling to an external field.

 \begin{figure}[!htb]
 \includegraphics[width=1\columnwidth]{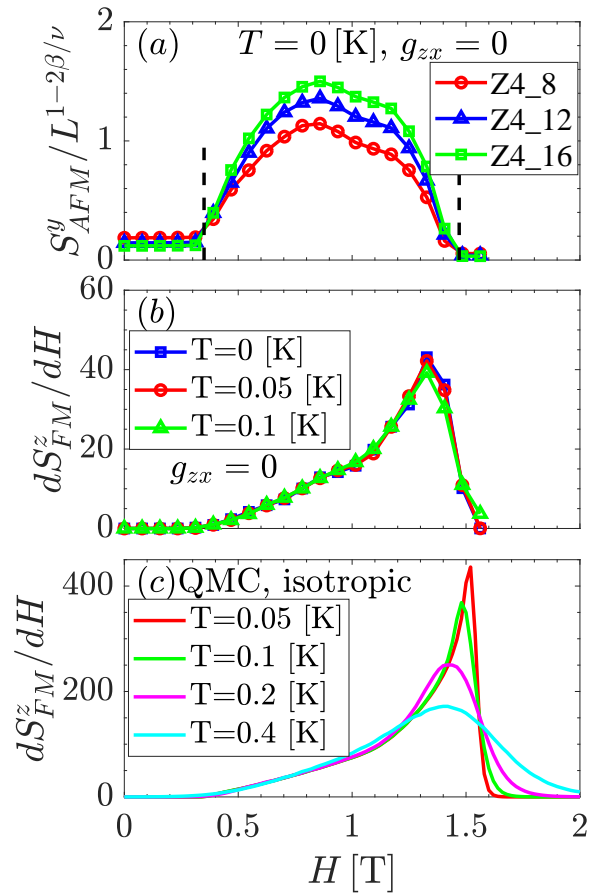}
 
 \caption{(a) Similar to Fig.~\ref{fig:DMRG_FSS}, but for the system without a tiny field in spin X direction, i.e. $g_{zx}=0$. Two crossings occur at $H_{c_1} \sim 0.4 \, \rm T$ and $H \sim 1.4 \, \rm T$. (b) Similar to Fig.~\ref{fig:SF_H}(b), but for the model with $g_{zx}=0$. The slope changes manifested in $dS_{FM}^z/dH$ versus $H$ suggest that a tiny $g_{zx}$ is not necessary to explain the similar feature in ultrasound velocity in the experiment \cite{Hester2019}.(c) QMC simulation results for a $16 \times 16$ honeycomb lattice with $512$ sites without any anisotropy, i.e. $\lambda=0$ and $g_{zx}=0$.}
 \label{isotropic}
 \end{figure}


\section{Discussion and Conclusions}
We have investigated the finite-temperature phase diagram and thermodynamics of the ``breathing" honeycomb lattice quantum dimer magnet in the parameter regime relevant to recent experiments on \ch{Yb2Si2O7}. We considered the effects of a small anisotropy in both the exchange coupling as well as the $g$-tensor, proposed as an explanation for the occurrence for two regimes inside antiferromagnetic regime in the Bose-Einstein condensation dome of \ch{Yb2Si2O7}~\cite{Flynn2021}. 

Our simulations employing the METTS technique yield close agreement with the experimentally observed data. By detecting maxima in the specific heat and performing finite-size scaling analysis of antiferromagnetic structure factors, we have mapped out the extent of the Bose-Einstein condensation dome which is found to closely track the experimentally observed data. 
Within the dome, two regimes have been distinguished in experiments by a change in the field dependence of the magnetization and the related ultrasound velocity measurements. This behavior is also found to be captured by the breathing honeycomb dimer model for which we observe a change of slope in the derivative of the ferromagnetic structure factor. Also we find close agreement when relating this quantity to the observed Bragg peak intensity and the related ultrasonic velocity measurements.

Our simulation data for specific heat Fig.~\ref{fig:C_T} fits the experimental data well for the full range of the magnetic field. The occurrence of a peak in the specific heat indicating a phase transition was pointed out in the experiment. However, this peak was only present in the lower-field regime of the BEC dome but absent in the higher-field regime, which was interpreted as another indication of two regimes. With the system sizes attainable using METTS we are at present unable to resolve sharp peaks, which would require simulating large fully two-dimensional geometries. Hence, the question whether or not a sharp peak in the specific heat is absent or present needs to be investigated further in future studies.

Moreover, we investigated to which extent anisotropies in the model are relevant to our findings. We confirm previous results that a small anisotropy in the $g$-tensor, $g_{xz} = g_{zz}/500$, leads to a phase transition at magnetic fields smaller than the saturation field at $T=0$ using DMRG. At temperatures above the anisotropy scale, however, this effect becomes negligible and we find that the actual phase transition is once again approximately concomitant with the saturation field. We conclude that the critical line in this model does not extend across the full temperature range of the Bose-Einstein condensation dome. The change in slope of the magnetization an the magnetic structure factor at finite-temperatures within the BEC dome is found to be a generic property occurring even in the fully isotropic case and is not related to a phase transition induced by the anisotropy. 

\section*{Acknowledgements}
We are grateful to Kate Ross and Gavin Hester for insightful discussions on the experiments on \ch{Yb2Si2O7} and for providing experimental data. We thank Rajiv Singh for helpful discussions about the interpretation of our results and comparisons to Ref.~\onlinecite{Flynn2021}.
The Flatiron Institute is a division of the Simons Foundation.

\bibliography{main.bib}

\end{document}